# Mechanical Cloak via Data-Driven Aperiodic Metamaterial Design

Liwei Wang[a,b], Jagannadh Boddapati[c], Ke Liu[c], Ping Zhu[a,*], Chiara Daraio[c,*], Wei Chen[b,*]

a. The State Key Laboratory of Mechanical System and Vibration, School of Mechanical Engineering, Shanghai Jiao Tong University, 800 Dongchuan Road, Shanghai 200240, China

b. Department of Mechanical Engineering, Northwestern University, 2145 Sheridan Road, Evanston, IL 60208, USA

c. Division of Engineering and Applied Science, California Institute of Technology,1200 E California Blvd, Pasadena, CA 91105, USA

* Ping Zhu

**Email:** pzhu@sjtu.edu.cn

* Chiara Daraio

**Email:** daraio@caltech.edu

* Wei Chen

**Email:** weichen@northwestern.edu

**Abstract**

Mechanical cloaks are materials engineered to manipulate the elastic response around objects to make them indistinguishable from their homogeneous surroundings. Typically, methods based on material-parameter transformations are used to design optical, thermal and electric cloaks. However, they are not applicable in designing mechanical cloaks, since continuum-mechanics equations are not form-invariant under general coordinate transformations.  As a result, existing design methods for mechanical cloaks have so far been limited to a narrow selection of voids with simple shapes. To address this challenge, we present a systematic, data-driven design approach to create mechanical cloaks composed of aperiodic metamaterials using a large pre-computed unit cell database. Our method is flexible to allow the design of cloaks with various boundary conditions, multiple loadings, different shapes and numbers of voids, and different homogeneous surroundings. It enables a concurrent optimization of both topology and properties distribution of the cloak. Compared to conventional fixed-shape solutions, this results in an overall better cloaking performance, and offers unparalleled versatility. Experimental measurements on 3D-printed structures further confirm the validity of the proposed approach. Our research illustrates the benefits of data-driven approaches in quickly responding to new design scenarios and resolving the computational challenge associated with multiscale designs of functional structures. It could be generalized to accommodate other applications that require heterogeneous property distribution, such as soft robots and implants design.

.

**Main Text**

**Introduction**

Metamaterials derive their properties mainly from the geometrical design of their microstructure besides their constituent materials(1-6) Metamaterials have been suggested for different functionalities in a wide range of applications, such as light-weight frames, biomimetic soft actuators, solar tracking systems, super-lenses, and invisibility cloaks.(7-14) As a representative application, cloaking materials could be used to conceal objects within a homogeneous surrounding, to prevent the detection of the objects with external physical fields, such as electromagnetic and mechanical fields.(7, 13-15) To achieve a cloaking effect, a cloak is designed around the object or void, with its material properties different from that of the surrounding homogeneous material (**Figure 1a**). A common approach for designing cloaks is through material-parameter transformation, by exploiting the form-invariance of governing equations under transformation to decide material properties within the cloak region. (8, 16-18) This approach has been successfully applied to, for example, cloaks for electromagnetic/optical waves,(19) static electricity,(20) and heat conduction. (21, 22) However, material-parameter transformation is not applicable to mechanical cloaks that conceal elastic responses, e.g., displacement field and elastic waves. This is because the theory of continuum mechanics is not form-invariant under general coordinate transformations—a prerequisite to use the material-parameter transformation. (23) Moreover, to characterize the mechanical response of materials, a four-rank elasticity tensor is required, which adds complications to the cloaking design problem, compared to designing the scalar material parameters in other physical problems. Mathematical derivation suggests that a perfect mechanical cloak requires microstructures to achieve special anisotropic Cosserat tensor distributions, which might not be physically achievable. (23, 24) As a result, limited progress has been made in designing mechanical cloaks. (24-27)

Although a theoretically perfect mechanical cloaking could not be achieved in reality, one can still obtain an approximate mechanical cloaking under some necessary assumptions and constraints, within a narrow tolerance. Along this line, to bypass the requirement of form-invariant equations in designing mechanical cloaks, a direct lattice point-transformation approach was developed for a special bi-mode lattice metamaterial. (24, 26, 28) Relying on a qualitative analog between electric conduction and mechanics, this approach heuristically applies spatial coordinate transformation on the lattice points, instead of effective material properties, within a prespecified cloak region (Figure 1a) for simple shapes of voids. In another study, to simplify the mechanics featured by the high-rank elasticity tensor, special parameterized pentamode materials, with nearly zero shear modulus, were exploited in realizing an approximate elasto-mechanical core-shell cloak for a circular void. (25) However, all of these methods are restricted to special types of parameterized lattices and can only cloak voids of simple shapes, i.e., circles and polygons. Moreover, the cloaking region must be pre-specified and cannot accommodate voids or solid objects with arbitrary shapes. In this work, we show that metamaterials with tessellated unit cells, selected from a pre-computed database, can be used to obtain a more general cloaking design approach.

We focus on elasto-static cloaks that retain the same elastic displacement field in the homogeneous material initially around the void (labeled "surrounding region" in Figure 1a). We propose a data-driven optimization method that capitalizes on a large and diverse precomputed metamaterial unit cell database. Our method concurrently optimizes the topology of the cloak and the distribution of metamaterial properties within the cloak. Based on this optimization, the method then assembles a physically fabricable structure made with different metamaterial unit cells through a tiling optimization process.(29)  By combining a large database with topology and properties optimization, this method can efficiently design cloaks to disguise multiple voids with complex shapes, in different homogeneous surroundings, under various boundary conditions, displaying superior flexibility compared to existing parametric design methods.

**Results**

As shown in Figure 1a, we focus on the design of a mechanical cloak, $\Omega_c$, around voids within the structure, that allows it to retain the displacement field in the surrounding region, $\Omega_s$, as if there were no voids. With this elastostatic cloak, voids could be intentionally introduced within a structure to realize special functions, e.g., to conceal underground tunnels, holes for wires and cavities to hide objects, without affecting the structural integrity and functionality of the original structure. To achieve this, as shown in Figure 1b, we consider metamaterials within the structure as homogenized continua, and then concurrently optimize the topology of the cloak, $\Omega_c$, as well as the spatial distribution of the effective properties. Mathematically, we optimize the independent entries of the stiffness tensor, within the cloak $\Omega_c$, to minimize the void-induced distortion of the displacement field in the surrounding region $\Omega_s$ (see SI for more details). For simplicity, we only consider unit cells with orthotropic symmetry in this study, whose stiffness tensor is calculated through energy-based homogenization and has four independent entries $C_{11}, C_{12}, C_{22}$ and $C_{33}$ for 2D design (in Voigt notation). (30, 31)  To ensure the fabricability of the optimized materials, the properties are constrained within the properties space of a large and dense pre-computed unit cell database, during the optimization process (Figure 1d and **S1**).  This pre-computed database is generated by combing unit-cell topology optimization and sequential stochastic shape perturbations algorithm to achieve shape and property diversities (see SI for more details and access for the dataset).  (32)

After the optimization, the cloak within the cloaking boundary is filled by an aperiodic tessellation of unit cells, selected to achieve the optimized properties distribution (Figure 1c shows a printed assembled structure). Specifically, a set of unit cells are selected from the database as candidates for each location in the cloak, with properties closest to the optimized local material properties. A tiling optimization is then performed to select the optimal unit cell from the candidate set in each location, to ensure good geometrical and mechanical compatibility between adjacent unit cells, as shown in Figure 1e. This tiling optimization is formulated as an energy-minimization problem on a grid-like graph and solved efficiently by a dual decomposition method (see SI for more details). (33, 34) After the tiling optimization, the assembled structure with a cloak is obtained (Figure 1c).

To demonstrate our approach, we first focus on the design of a cloak for a circular void, fixing the shape and size of the cloak. As shown in **Figure 2a**, the reference structure (without voids and cloak) is composed of $30 \times 30$ periodically tessellated base cells, each of size $5 \text{ mm} \times 5 \text{ mm}$, whose constituent material has Young's modulus $E = 1.20 \text{ GPa}$ and Poisson's ratio $v = 0.35$. To facilitate displacement tracking in experimental measurement, a simple four-rod cubic lattice is chosen as the base cell (shown in the inset of **Figure 2a**). Its homogenized elastic properties, calculated through energy-based homogenization, are: $C_{11} = 171.55 \text{ MPa}, C_{12} = 69.68 \text{ MPa}, C_{22} = 171.55 \text{ MPa}$, and $C_{33} = 62.07 \text{ MPa}$. Each unit cell is discretized into a $50 \times 50$ plane stress Q4 element. Finite element simulations are then performed to calculate the displacement field of the reference structure under different boundary conditions (see SI for more details). We first model the response of the structure to a compressive load $u_{bc} = 0.5 \text{ mm}$ ($0.67\%$ average strain) in the horizontal direction, while keeping the top and bottom edges free (we refer to this configuration as 'displacement-free' boundary condition). The response of the material is assumed to remain linear elastic during the loading process. The resultant displacement fields in the homogeneous material, in both the *x* and *y* directions, are shown in Figure 2b and 2c, exhibiting a uniform transition from one end to the other. The presence of a circular void at the center of the structure (aka., in the 'voided structure', Figure 2f) causes a large distortion of the displacement field (Figure 2g and 2h), with significant shrinkage of the circular void. To control this distortion, we introduce a ring-shaped cloak, $\Omega_c$, around the void, whose outer radius is twice the radius of the central void, as commonly used in the literature where the cloaking boundary is predefined. (24) Later, we will also present results when the cloaking boundary is not predefined but instead designed. The homogeneous medium outsides the cloak in the structure, filled by periodic base cells as in the reference structure, is denoted as $\Omega_s$. Using the proposed method, the cloak $\Omega_c$ is filled with aperiodic unit cells selected from the database to minimize the void-induced distortion of the displacement fields within the region $\Omega_s$. We refer to this structure with a cloak surrounding the void as the 'cloaked structure', which is shown in Figure 2k. With our tiling optimization, neighboring unit cells retain good connections between each other, even though their properties and geometry vary. It is noted that the proposed optimization method allocates unit cells in a way that gradually increases the stiffness from the outer boundary to the inner one of the cloak $\Omega_c$ (Figure 2k and 2p). This gradient of stiffness compensates for the lack of structural cohesion due to the presence of the void, which is in line with existing mechanical cloak designs. Successful cloaking is evident when comparing the displacement fields of the reference structure, in Figure 2b and 2c, with the one in $\Omega_s$ of the cloaked structure, in Figure 2l and 2m. The large distortion of the central void is mitigated by the optimized cloak, displaying a gradual and uniform transition of displacement field along both *x*- and *y*-directions in the surrounding region $\Omega_s$ as that of the reference structure (i.e., without the void).

As a quantitative measure for the distortion of the displacement field, we opt for the commonly used relative displacement difference Δ formulated as

$$\Delta = \frac{\sqrt{\Sigma_{\Omega_s}(\vec{u}_i - \vec{u}_{0,i})^2}}{\sqrt{\Sigma_{\Omega_s}(\vec{u}_{0,i})^2}},\qquad(1)$$

where $\vec{u}_i$ represents the nodal displacements of the finite elements in the cloaked (Figure 2l and 2m) or voided structure (Figure 2g and 2h), and $\vec{u}_{0,i}$ represents the nodal displacements of the reference structure (Figure 2b and 2c). The summation operator performs the sum over all the nodes within the surrounding region $\Omega_s$. The lower the relative displacement difference Δ, the smaller the overall distortion of the displacement field from the reference state. With the designed cloak, the relative difference calculated from the numerical simulations is reduced from 17.2% in the voided structure to 4% in the cloaked structure, demonstrating a good cloaking performance.

To further validate the findings, we perform experiments on 3D printed structures, fabricated to reproduce (i) the reference structure, (ii) the voided structure, and (iii) the cloaked structure (Figure 2q). Considering the symmetry of the structure and the camera's field of view, we measure displacements from the upper left corner of each structure (Figure 2d). Equation 1 is used to calculate the relative displacement difference Δ in the experiment by inserting the measured displacements of the lattice points (centers of base cells) within the region $\Omega_s$. Overall, the simulated and experimental displacements match well in terms of both displacement distribution and the relative displacement difference (compare, e.g., Figure 2a,b with Figure 2d,e, or Figure 2g,h with Figure 2i,j).

The results shown in Figure 2 are obtained under the displacement-free boundary condition. Nevertheless, our method is general and valid for various other boundary conditions. We demonstrate this by performing additional optimizations and numerical simulations with pressure-free, pressure-sliding, dilating, and shearing boundary conditions (**Figure 3a-f, Video S1**). For the pressure-free boundary condition, we apply constant compressive pressure $p_{bc} = 95.67$ KPa on the left and right edges, to keep average strain around 1% and ensure that the structure remains in the linear elastic deformation range. The proposed optimization method generates a cloaked structure shown in Figure 3c, with a similar stiffness gradient as observed in the design obtained for the previous displacement-free boundary condition (Figure 2k, p, q). As shown in the first two bar groups in Figure 3b, the relative displacement difference of the voided structure is much higher for the pressure-free boundary condition (102.3%) than that of the previous displacement-free boundary condition (17.2%). This is because, in contrast to the displacement-free boundary condition, the pressure-free boundary condition allows nonuniform displacements on the left and right boundaries, which encourages the shrinkage of the void after the loading (**Figure S5**). Nevertheless, the cloaked structure obtained can still effectively reduce the distortion caused by the void to a $\Delta = 12.8\%$. Adding an extra sliding boundary condition on the top and bottom in the pressure-free boundary condition leads to the pressure-sliding boundary condition. Under this condition, the constraints imposed by the sliding support compress the structure against Poisson's-effect-induced expansion in the vertical direction. As a result, compared with the previous single-directional loading cases (displacement-free and pressure-free), the joint effects of the support reactions and imposed pressure enable the void-induced distortion to cover a larger area in the structure (**Figure S6**). Therefore, the optimized cloak contains more unit cells with high stiffness (Figure 3d and S6), to balance the strains and suppress the shrinkage of the void. The expansion of the high stiffness region is even more obvious under the dilating boundary condition (Figure 3e). This is because, under this boundary condition, the affected area of the distortion in the voided structure is even larger with the uniformly distributed stretching force exerted from all four edges (**Figure S7**). Similarly, for the shearing boundary condition, large void-induced distortions of the displacement field mainly distribute in the right half of the structure (**Figure S8**), corresponding to regions with stiffer unit cells in the cloak (Figure 3f). Nevertheless, neighboring unit cells in the cloak remain well-connected for these three boundary conditions with our tiling optimization, achieving excellent cloaking performance with low relative errors (< 9%), as shown in the third through fifth bar groups, in Figure 3b. Note that all the cloak designs shown in this study remain effective for other magnitudes of compressive pressure or displacement imposed on the boundary, as long as the deformation remains in the regime of linear elasticity. This is because, in linear elasticity, the relative displacement difference, Δ, does not depend on the absolute magnitude of the loading.

While most existing design methods for mechanical cloaks confine the base cell in the surrounding region to bi-mode or simple parameterized lattice, the proposed method can systematically accommodate various free-formed base cells. As an example, we use a different type of unit cell with more complex geometry as the base cell and consider a circular void under the pressure-free boundary condition (Figure 3g). The homogenized stiffness tensor of this base cell yields: $C_{11} = 123.02 \text{MPa}, C_{12} = 48.02 \text{ MPa}, C_{22} = 123.02 \text{ MPa}$, and $C_{33} = 13.50 \text{ MPa}$. As shown by the simulated displacement field (**Figure S9, Video S2**) and the large relative displacement difference of the voided structure (Figure 3b), this type of compliant base cell leads to a significant amount of distortion of the displacement field. Yet, the proposed method can still effectively reduce the relative error from 139.1% to 19.6% with the optimized cloak. Moreover, it should be noted that the proposed method can also design the same cloak to accommodate various loading cases by aggregating individual cloaking performance into an overall objective function. For example, we obtain a cloak that achieves excellent cloaking performance (Δ < 15%) for constant distributed loading imposed from any angle (**Figure S10** and **S11**), by using a p-norm-based aggregated objective function in the optimization (see SI for more details).

Unlike design methods based on the material-parameter transform or direct lattice transform, the proposed method can also be applied to create cloaks around voids with arbitrary shapes (Figure 3h-k, **S12** and **Video S3**), due to either functional or aesthetic needs. We demonstrate this capability by designing cloaks for different voids, applying the same pressure-free boundary condition and predefined cloak topology, as in Figure 2. Similar to our previous designs for a circular void, the stiffness of the unit cells within these optimized cloaks (Figure 3h-k) exhibit a decreasing trend starting from the edge of the void to the outer boundary of the cloak. From Figure 3b and S12, it can be noted that the various optimized cloaks reduce distortion of the displacement fields induced by the voids of various shapes, keeping relative displacement difference at a considerably low level (6.8%~13.5%). It should be noted that, while the circular void in Figure 3c could geometrically enclose voids of various shapes in Figure 3h-j, the solution is less optimal or efficient compared to the cloaking design optimized specifically for a desired shape of voids.

In all results discussed so far, a predefined cloaking boundary of the cloak is used and remains unchanged during the design process. This is a requirement similar to most existing methods. (24-26, 28) We will present next that our proposed method can also simultaneously design the topology, i.e., free-formed geometry that allows topological changes, of the cloaking region $\Omega_c$ and the property distribution within it, adapting to different geometries of voids. To demonstrate the benefits of not predefining the cloaking boundary, based on the Mickey-shaped void, we limit the area of the cloak to be equal to that of the previously used circular one, while concurrently optimizing the topology and its property distribution to minimize the relative displacement difference (**Video S4**). The resulting cloaked structure and its displacement field are shown in Figure 3l and S12, respectively. We observe that the cloaking boundary conforms to the contour of the Mickey-shaped void, expanding in the upper half and shrinking in the lower half. As shown in Figure 3b, while the relative displacement differences are similar for the voided structures with the predefined (88.3%) and optimized topologies of cloaks (86.7%), the cloak with optimized topology can achieve a much smaller relative error (8.1%) than the cloak with predefined topology (13.5%).

To further demonstrate the versatility of the proposed method, we design a cloak for two voids arranged in a butterfly shape, subjected to the displacement-free boundary condition, as shown in **Figure 4**. The specialty of this example lies in that there are two neighboring irregular voids (Figure 4f), which cannot be decomposed into simple ellipses or squares. The existence of multiple voids and the mechanical interaction between them poses an additional challenge to the adaptive design of mechanical cloaks. As shown in Figure 4g-4j and **Video S5**, both the simulated and experimental results suggest that the butterfly-shaped voids distort the displacement field significantly, especially along the *y*-direction. The cloaked structure designed by the proposed method successfully reduced the relative difference Δ nearly by half for both calculated and experimental results, leading to a displacement field (Figure 4l-4o) close enough to that of the reference structure (Figure 4b-4e).

**Conclusion**

In this study, we have developed a data-driven method for the design of mechanical cloaks, powered by topology optimization. This method first optimizes the shape and topology of the cloak and the material property distributions within the cloak, and then selects optimal unit cells from a precomputed database of material microstructures to fill the cloak through an efficient optimization-based tiling process. Using multiple examples with various numbers and shapes of voids, boundary conditions, and base cells, we demonstrate that our method achieves excellent performance in mechanical cloaking, verified both numerically and experimentally. Compared to existing approaches that are only applicable to a fixed but restricted topology of cloaks, our method is capable to concurrently design the topology and property distribution of the cloak, balancing both efficiency and versatility. We observe that the optimized topology of the cloak generally conforms to the contour of different geometries of voids, providing a better cloaking performance compared to cloak designs with fixed topology. The data-driven approach is powerful in terms of its capability to swiftly respond to new design scenarios with limited computational cost, by taking advantage of the rich unit-cell database that is readily available. In the preparation of the databases, we can also easily take into account manufacturability and other resource restrictions.

**Methods**

The design of each unit cell in the database is pixelated and represented by a $50 \times 50$ binary matrix. We follow the common practice in multiscale topology optimization to use a corresponding $50 \times 50$ four-node quadrilateral plane stress finite element mesh for each unit cell in the full structure. Because the maximal strain of the assembled structure is around 1%~2%, the finite element analysis is carried out in MATLAB and ABAQUS under the linear elasticity and plane stress assumptions. The good agreement between the numerical calculation and the experimental measurement validates the effectiveness of the calculation method.

We fabricated all the structures tested using a Stratasys Connex Objet 500 3D printer. We subjected the structures to compression using an Instron E3000 machine. While the structures were being loaded, we captured a sequence of images using a Nikon D750 camera mounted with a Nikkor 200mm f/4D IF-ED lens. Digital image correlation on the captured images was performed with a global DIC code designed specifically for two-dimensional materials with microstructures. (35)

More details on the design methods and experiments can be found in the supplementary information.

**Acknowledgments**

We are grateful for support from the NSF CSSI program (Grant No. OAC 1835782) and Center for Hierarchical Materials Design (ChiMaD NIST 70NANB19H005). Liwei Wang acknowledges support from the Zhiyuan Honors Program for Graduate Students of Shanghai Jiao Tong University for his predoctoral visiting study at Northwestern University.

**References**

1. M. Wegener, Metamaterials beyond optics. *Science* **342**, 939-940 (2013).
2. N. I. Zheludev, The road ahead for metamaterials. *Science* **328**, 582-583 (2010).
3. X. Yu, J. Zhou, H. Liang, Z. Jiang, L. Wu, Mechanical metamaterials associated with stiffness, rigidity and compressibility: A brief review. *Progress in Materials Science* **94**, 114-173 (2018).
4. M. Kadic, G. W. Milton, M. van Hecke, M. Wegener, 3D metamaterials. *Nature Reviews Physics* **1**, 198-210 (2019).
5. D. R. Reid *et al.*, Auxetic metamaterials from disordered networks. *Proceedings of the National Academy of Sciences* **115**, E1384-E1390 (2018).
6. T. S. Lumpe, T. Stankovic, Exploring the property space of periodic cellular structures based on crystal networks. *Proceedings of the National Academy of Sciences* **118** (2021).
7. U. Leonhardt, Optical conformal mapping. *science* **312**, 1777-1780 (2006).
8. A. Greenleaf, Y. Kurylev, M. Lassas, U. Leonhardt, G. Uhlmann, Cloaked electromagnetic, acoustic, and quantum amplifiers via transformation optics. *Proceedings of the National Academy of Sciences* **109**, 10169-10174 (2012).
9. B. Jenett *et al.*, Discretely assembled mechanical metamaterials. *Science advances* **6**, eabc9943 (2020).
10. A. Rafsanjani, Y. Zhang, B. Liu, S. M. Rubinstein, K. Bertoldi, Kirigami skins make a simple soft actuator crawl. *Science Robotics* **3** (2018).
11. A. Lamoureux, K. Lee, M. Shlian, S. R. Forrest, M. Shtein, Dynamic kirigami structures for integrated solar tracking. *Nature communications* **6**, 1-6 (2015).

12. N. Kaina, F. Lemoult, M. Fink, G. Lerosey, Negative refractive index and acoustic superlens from multiple scattering in single negative metamaterials. *Nature* **525**, 77-81 (2015).
13. Y. Yang *et al.*, Full‐polarization 3D metasurface cloak with preserved amplitude and phase. *Advanced Materials* **28**, 6866-6871 (2016).
14. S. Narayana, Y. Sato, DC magnetic cloak. *Advanced Materials* **24**, 71-74 (2012).
15. J. B. Pendry, D. Schurig, D. R. Smith, Controlling electromagnetic fields. *science* **312**, 1780-1782 (2006).
16. L. Xu, H. Chen, Conformal transformation optics. *Nature Photonics* **9**, 15-23 (2015).
17. H. Chen, C. T. Chan, P. Sheng, Transformation optics and metamaterials. *Nature materials* **9**, 387-396 (2010).
18. U. Leonhardt, Notes on conformal invisibility devices. *New Journal of Physics* **8**, 118 (2006).
19. W. Cai, U. K. Chettiar, A. V. Kildishev, V. M. Shalaev, Optical cloaking with metamaterials. *Nature photonics* **1**, 224-227 (2007).
20. F. Yang, Z. L. Mei, T. Y. Jin, T. J. Cui, DC electric invisibility cloak. *Physical review letters* **109**, 053902 (2012).
21. R. Schittny, M. Kadic, S. Guenneau, M. Wegener, Experiments on transformation thermodynamics: molding the flow of heat. *Physical review letters* **110**, 195901 (2013).
22. T. Chen, C.-N. Weng, Y.-L. Tsai, Materials with constant anisotropic conductivity as a thermal cloak or concentrator. *Journal of Applied Physics* **117**, 054904 (2015).
23. G. W. Milton, M. Briane, J. R. Willis, On cloaking for elasticity and physical equations with a transformation invariant form. *New Journal of Physics* **8**, 248 (2006).
24. T. Bückmann, M. Kadic, R. Schittny, M. Wegener, Mechanical cloak design by direct lattice transformation. *Proceedings of the National Academy of Sciences* **112**, 4930-4934 (2015).
25. T. Bückmann, M. Thiel, M. Kadic, R. Schittny, M. Wegener, An elasto-mechanical unfeelability cloak made of pentamode metamaterials. *Nature communications* **5**, 1-6 (2014).
26. L. Hai, Q. Zhao, Y. Meng, Unfeelable mechanical cloak based on proportional parameter transform in bimode structures. *Advanced Functional Materials* **28**, 1801473 (2018).
27. X. Xu *et al.*, Physical realization of elastic cloaking with a polar material. *Physical review letters* **124**, 114301 (2020).
28. M. Kadic *et al.*, Elastodynamic behavior of mechanical cloaks designed by direct lattice transformations. *Wave Motion* **92**, 102419 (2020).
29. L. Wang *et al.*, Deep generative modeling for mechanistic-based learning and design of metamaterial systems. *Computer Methods in Applied Mechanics and Engineering* **372**, 113377 (2020).
30. Z. Hashin, Analysis of Composite Materials—A Survey. *Journal of Applied Mechanics* **50**, 481 (1983).
31. L. Xia, P. Breitkopf, Design of materials using topology optimization and energy-based homogenization approach in Matlab. *Structural and multidisciplinary optimization* **52**, 1229-1241 (2015).
32. L. Wang, Y.-C. Chan, Z. Liu, P. Zhu, W. Chen, Data-driven metamaterial design with Laplace-Beltrami spectrum as "shape-DNA". *Structural and Multidisciplinary Optimization*, 1-16 (2020).

33. C. Wang, N. Komodakis, N. Paragios, Markov random field modeling, inference & learning in computer vision & image understanding: A survey. *Computer Vision and Image Understanding* **117**, 1610-1627 (2013).
34. N. Komodakis, N. Paragios, G. Tziritas, MRF energy minimization and beyond via dual decomposition. *IEEE transactions on pattern analysis and machine intelligence* **33**, 531-552 (2010).
35. F. Agnelli, P. Margerit, P. Celli, C. Daraio, A. Constantinescu, Systematic two-scale image analysis of extreme deformations in soft architectured sheets. *International Journal of Mechanical Sciences* **194**, 106205 (2021).

**Figures and Tables**

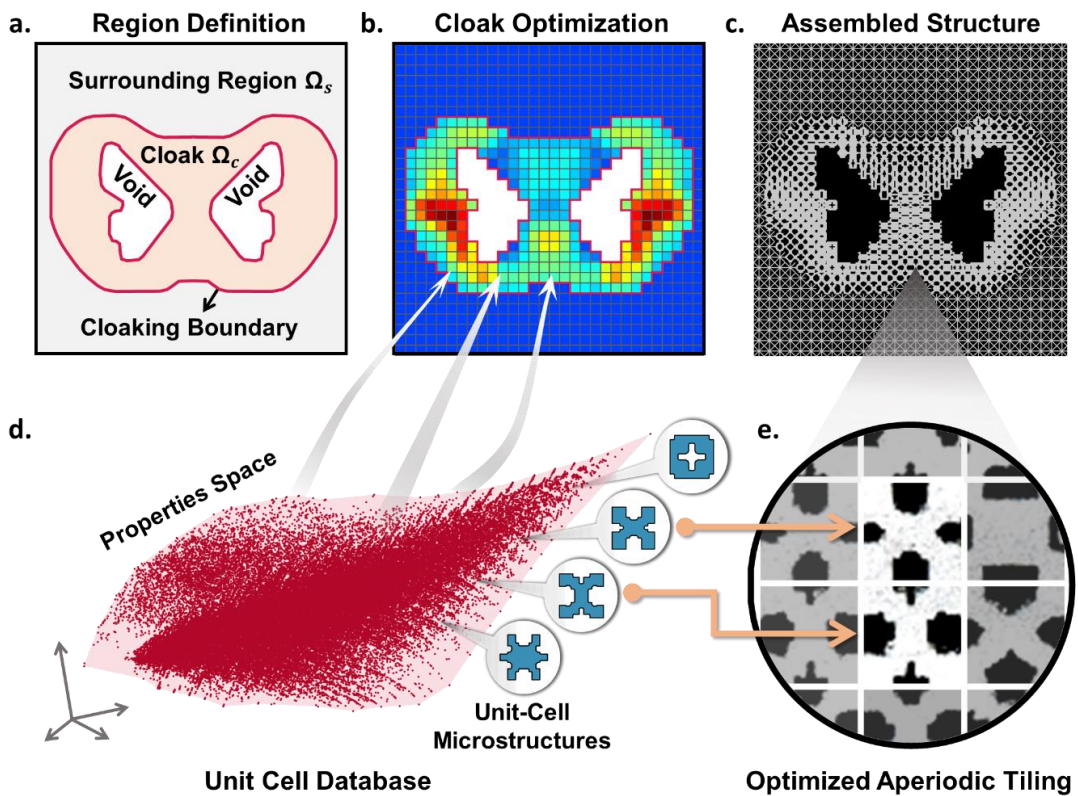

**Figure 1.** Schematic diagram of the data-driven design of mechanical cloaks. Panel (a) shows the region definition adopted in this study with voids, cloak and surrounding region colored in white, orange and grey, respectively. Panel (b) demonstrates cloak optimization result, with both its topology (enclosed by red lines) and properties distribution (marked with gradient color) concurrently optimized. Panel (c) shows the assembled structure (3D printed for validation) corresponding to the cloak optimization. Panel (d) exhibits the properties space and representative unit-cell microstructures of the precomputed unit cell database. It provides properties space to cloak optimization and candidate unit cells to achieve (e) optimized aperiodic unit-cell tiling in the 3D printed assembled structure.

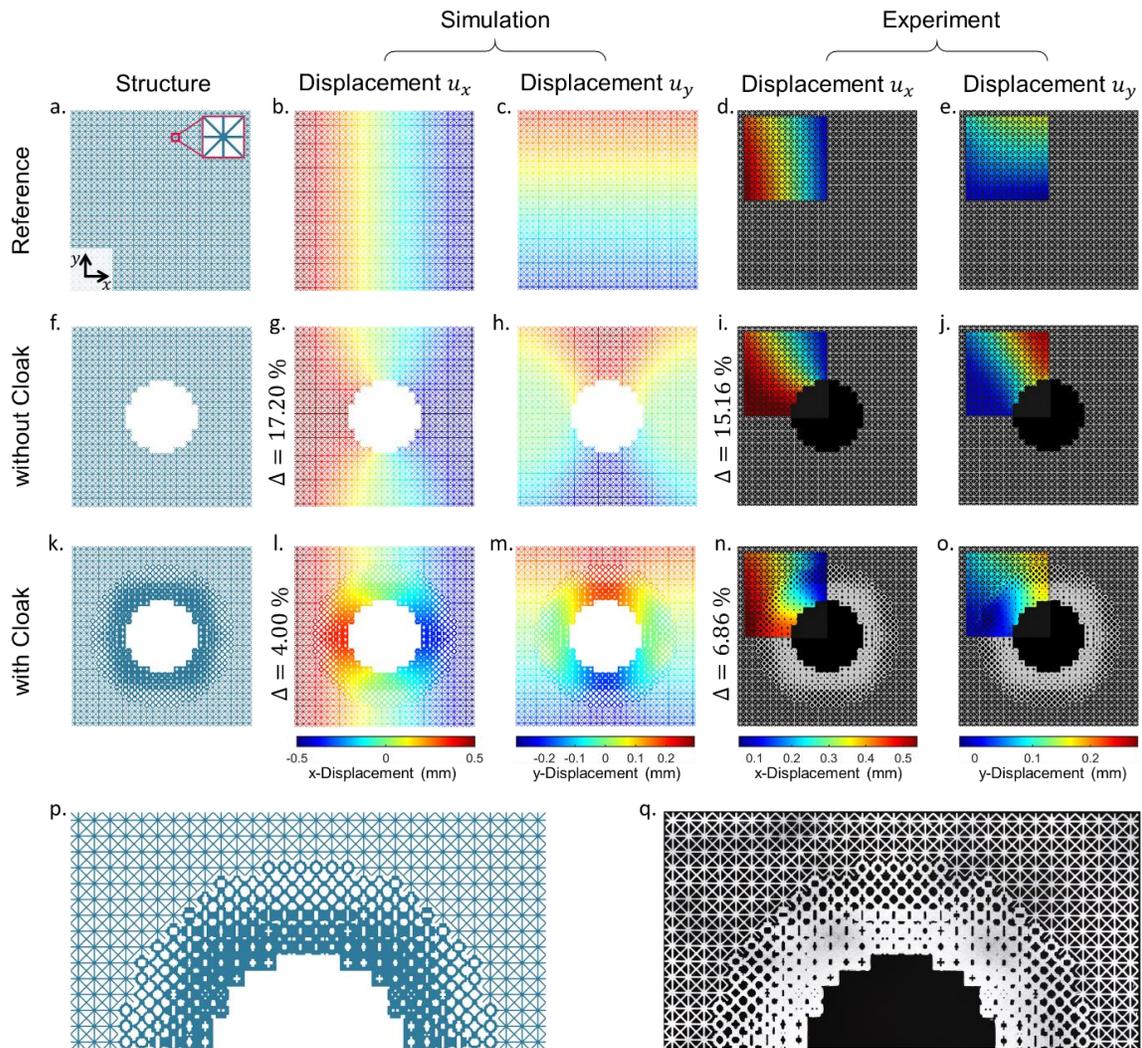

**Figure 2.** Comparison of the numerical and experimental displacement fields obtained for the reference (panels a-e), voided (f-j), and cloaked structures (k-o). Panels (a), (f), (k) show the geometry of the structures. Panels (p) and (q) show enlarged images of the upper half of the computationally designed, and 3D physically printed cloaks, respectively. Inset in panel (a) shows the base cell of the reference structure. Constant horizontal displacements are imposed on the left and right boundaries while keeping the other boundaries free.

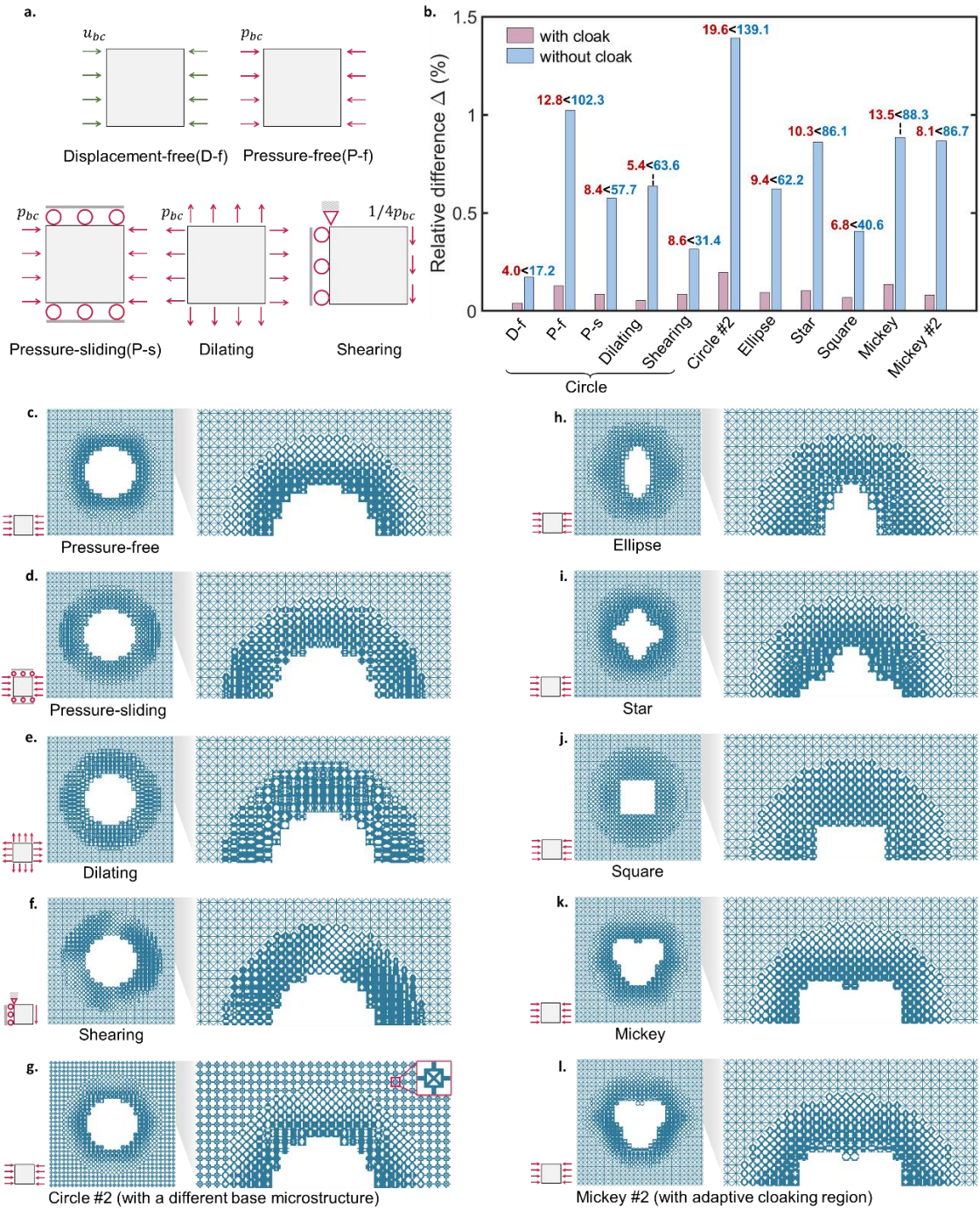

**Figure 3.** Design results for different boundary conditions, shapes of the void and a different type of base cell. (a) Different boundary conditions analyzed, (b) bar graph for the relative displacement differences of different voided (values colored in red) and cloaked structures (values colored in blue), cloaked structures for a circular void under (c) pressure-free boundary condition, (d) pressure-sliding boundary condition, (e) dilating boundary condition and (f) shearing boundary condition, (g) cloaked structure for a circular void under pressure-free boundary condition with the second type of base cell shown in the inset of the enlarged image, (h)~(k) cloaked structures for different shapes of voids under the same pressure-free boundary

condition. (l) cloaked structure for a Mickey-shaped void under the pressure-free boundary condition with optimized topology of the cloak. Enlarged images of the upper half of cloaked structures are shown on the right in panels (c)~(l).

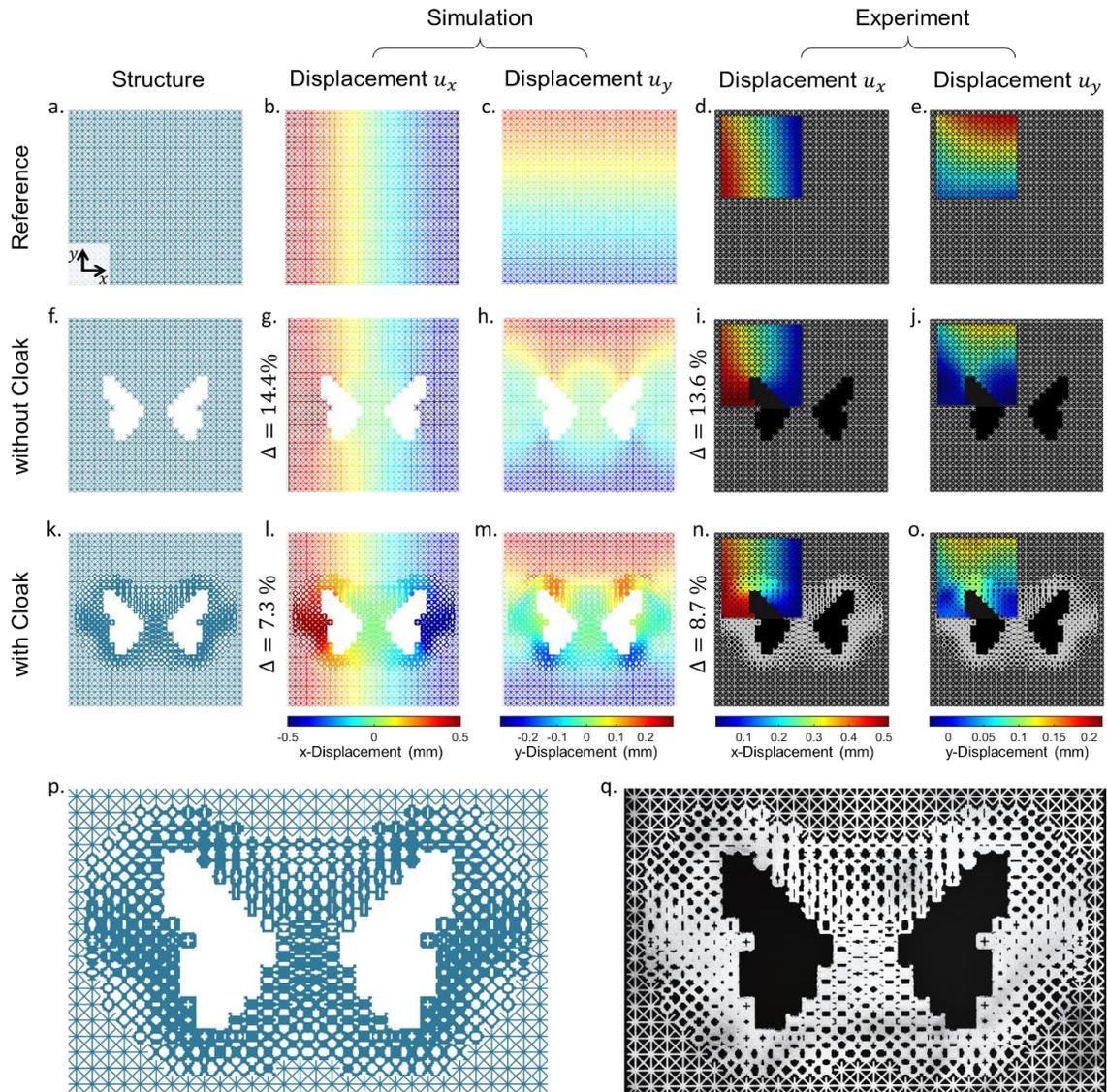

**Figure 4.** Geometries, calculated and experimentally measured displacement fields for reference, voided, and cloaked structures as that in Figure 2, but with butterfly-shaped voids and optimized topology of the cloak. Panels (p) and (q) show enlarged images of the designed and 3D printed cloaks, respectively.

**Supplementary Information Text**

**Unit Cell Database Generation**

We construct a large database with diverse unit-cell microstructures and pre-computed properties following our previously proposed method with a combination of topology optimization and an iterative stochastic shape perturbation.(1, 2) This database is available on our website (https://ideal.mech.northwestern.edu/research/software/). We focus on the orthotropic unit cells, with only four independent elements for their corresponding, homogenized stiffness tensor, i.e., $C_{11}, C_{12}, C_{22}$ and $C_{33}$ (in Voigt notation). The constituent material has Young's modulus $E = 1.20$ GPa and Poisson's ratio $\nu = 0.35$. To cover a wide range of properties, we first perform SIMP-based TO to find a corresponding pixelated unit-cell design of size $5 \text{ mm} \times 5 \text{ mm}$, for each uniformly sampled target stiffness matrix.(3) In each iteration of TO, the energy-based homogenization method is used to calculate the effective (homogenized) stiffness matrix.(4) This TO design process generates 1400 unit cells, each represented by a 50×50 binary matrix with zero and one corresponding to void and solid, respectively. With these unit cells as initial seeds, an iterative stochastic shape perturbation algorithm is employed to perturb unit-cell geometries that have extreme (close to the boundary of the properties space) or uncommon properties (with a small number of neighbors in the properties space) in the database.(1) Unit cells with isolated pixels, checkerboard patterns, or features smaller than the predefined minimum length scale (0.5 mm) are filtered out to ensure the manufacturing feasibility. By performing the selection and perturbation process for 100 iterations using parallel computing, we create a large database with 90245 unit cells in our study with the property space shown in **Figure S1**. Note that, while a smaller database might be enough for some special cloak designs, we generate this large database to accommodate general design cases by providing a large range of properties and diverse candidates for the later optimization. In this way, the database generation only needs to be performed once in an offline manner, and the generated database can be readily applied to various design cases.

**Concurrent Topology and Properties Optimization of the Mechanical Cloak**

We propose a design method that could concurrently optimize the topology and properties of the cloak for a better cloaking performance. The structure is divided into a matrix of equal-size square subdomains ($30 \times 30$ in this study). Each subdomain outside the void is assumed to be filled by either the given base cell (in the homogeneous surrounding region) or a unit cell selected from the database (in the cloak). In the concurrent optimization, each subdomain is modeled as a single Q4 element in the optimization process, with the homogenized stiffness tensor of the corresponding unit cell as its elemental material properties. As shown in **Figure S2a**, we define a topology design variable $x^e \in \{0,1\}$ for each subdomain with zero and one to indicate the surrounding region $\Omega_s$ and the cloak $\Omega_c$, respectively. A normalized property vector $z^e \in [0,1]$ is defined for each subdomain in the design region $\Omega_s \cup \Omega_c$ so that the elemental properties at that subdomain $\widetilde{C}^e$ can be formulated as:

$$\widetilde{C}^e = (\widetilde{C}^+ - \widetilde{C}^-) \circ z^e + \widetilde{C}^-, \tag{S1}$$

where $\widetilde{C} = [\tilde{C}_1, \tilde{C}_2, \tilde{C}_3, \tilde{C}_4] = [C_{11}, C_{12}, C_{22}, C_{33}]^T$ is a vector to denote independent entries of the homogenized stiffness matrix, $\widetilde{C}^+$ and $\widetilde{C}^-$ are the upper and lower bounds of $\widetilde{C}$ for unit cells in the constructed database. To correlate the material properties with the topology design variable $x^e$ for concurrent optimization, $z^e$ is further given as:

$$z^e = x^e \cdot y^e + (1 - x^e) \cdot z_0, \tag{S2}$$

where $z_0$ is the corresponding normalized property vector for the base unit cell and $y^e \in [0,1]$ is the property design vector, as shown in Figure S2b and S2c. For subdomains within the cloak, i.e., $x^e=1$, the second term in Equation S2 diminishes so that $z^e = y^e$, and $\widetilde{C}^e$ in Equation S1 is fully determined by the property design vector $y^e$. In contrast, for subdomains in the $\Omega_s$, i.e., $x^e=0$, the first term in Equation S2 equals zeros so that $z^e = z_0$, and the property of the base cell is assigned to these subdomains.

To enable the use of gradient-based optimizers for higher efficiency, we relax the integral variable $x^e$ to be continuous on [0,1] by using the following projection scheme:

$$\tilde{x}^e = H(x^e|\beta,\eta) = \frac{tanh(\beta\eta)+tanh(\beta(x^e-\eta))}{tanh(\beta\eta)+tanh(\beta(1-\eta))} \quad \text{with } x^e \in [0,1], \tag{S3}$$

where $\tilde{x}^e$ is the projected topology design variable and $H(\cdot|\beta,\eta)$ is an approximated Heaviside function with parameters $\beta$ and $\eta$ to control the threshold and sharpness of the projection, respectively, as shown in **Figure S3**. In this study, we set $\eta = 0.5$ and gradually increase $\beta$ during the optimization to drive $\tilde{x}^e$ to converge to a binary value. To encourage this convergence, a penalization is further imposed on the normalized property vector $z^e$ for $\tilde{x}^e$ with intermediate value:

$$z^e = (\tilde{x}^e)^P \cdot y^e + [1 - (\tilde{x}^e)^P] \cdot z_0, \tag{S4}$$

where $P$ is a parameter to control the level of penalization and set to be 3 in this study. Since the cloak $\Omega_c$ and surrounding region $\Omega_s$ are changing iteratively during the design process, we need to define a generalized relative difference Δ of the displacement field for the iterative update of the design parameters. To achieve this, instead of explicitly selecting nodal displacement vectors of nodes in the surrounding region $\Omega_s$ for the summation operation in Equation 1, we assign weights for all the nodes in both the cloak $\Omega_c$ and surrounding region $\Omega_s$, represented by a weight vector $\gamma$. For the $i^{th}$ node, subdomains around this node (within $\Omega_s \cup \Omega_c$) are ordered anti-clockwise, starting from the lower-left direction. Value 1-$\tilde{x}^e$ of the first subdomain $e$ in this ordered list is assigned to entries $\gamma_{2i-1}$ and $\gamma_{2i}$ of the weight vector $\gamma$, serving as the weights for x- and y-displacements of this node. In this way, displacements of nodes in $\Omega_s$ with be automatically assigned with larger weights while that of nodes within the cloak $\Omega_c$, whose displacement vectors have little contribution to the relative displacement difference, will get smaller weights during the optimization. With these, the optimization of the mechanical cloak can now be formulated as

$$\min_{x^e, y^e} \mathcal{F} = \|\gamma \odot (u - u_0)\|_2^2 / \|\gamma \odot u_0\|_2^2$$
$$s.t. K(x^e, y^e)u = f,$$
$$V_* - \frac{1}{N}\sum_{e=1}^{N} \tilde{x}^e = 0, \tag{S5}$$
$$\varphi_e(y^e) \geq 0$$
$$0 \leq x^e \leq 1, \ 0 \leq y_i^e \leq 1, i = 1, \dots, 4$$

Where the objective function $\mathcal{F}$ is the square of relative displacement difference Δ, $\odot$ represent the element-wise product, $u$ and $f$ are the displacement and load vectors respectively, $u_0$ is the original displacement vector in the reference structure, $K$ is the global stiffness matrix depending on $x^e$ and $y^e$ of subdomain $e$, $V_*$ is the constraint on the ratios of cloaking area $\Omega_c$ with respect to the whole design region $\Omega_s \cup \Omega_c$, $N$ is the number of subdomains within the design region, and $\varphi$ is the inequality constraint to force feasible properties.

To formulate the property constraint $\varphi$, the value of signed L2 distance to the boundary is calculated for each node on a Cartesian grid enclosing the property space spanned by $C_{11}, C_{12}, C_{22}$ and $C_{33}$, with positive and negative values to indicate regions inside and outside the boundary, respectively. The signed L2 distance field and its partial derivatives within the grid can then be efficiently estimated by interpolation. By using this signed L2 distance field as the constraint function $\varphi$, we can ensure the optimized properties are achievable with unit cells in the constructed database. However, this unit-cell constraint will result in an immense number of constraints, making the iterative optimization process extremely time-consuming. To address this issue, we first project the value of constraint $\varphi$ of each subdomain into the region of $[-1,1]$ through approximated Heaviside projection:

$$S(\varphi) = \tfrac{1}{2}(\tanh(-\theta\varphi) + 1), \tag{S6}$$

where $\theta$ is a parameter to control the steepness of the projection and set to be 64 in this study.(5) With this projection, an equivalent optimization problem is then obtained to enable a more efficient solving process by aggregating local constraints in numerous subdomains into a single global constraint:

$$\begin{aligned}&\min_{x^e,y^e} \mathcal{F} = \|\gamma \odot (u - u_0)\|_2^2 / \|\gamma \odot u_0\|_2^2 \\ &s.t. \, K(x^e, y^e)u = f, \\ &V_* - \tfrac{1}{N}\sum_{e=1}^{N} \tilde{x}^e = 0, \\ &\tfrac{1}{N}\sum_{e=1}^{N} S(\varphi_e(y^e)) \leq \tfrac{1}{N} \\ &0 \leq x^e \leq 1, \; 0 \leq y_i^e \leq 1, i = 1, \dots, 4\end{aligned} \tag{S7}$$

The optimized solution of this optimization problem is obtained through the method of moving asymptotes (MMA), representing the topology ($x^e$) and the associated property distribution ($y^e$) of the cloak.(6) The optimization process will terminate when the number of iterations reaches 1000 or the maximal change of design variables in two consecutive iterations is lower than 0.001. Cloaked structures with optimized topologies of cloaks in Figure 3l and Figure 4 were designed through this method. For other cloak designs in this study with a fixed and predefined topology of the cloak, we used $y^e$ of subdomains within the cloak as the only design variables. Correspondingly, values of $\tilde{x}^e$ and $x^e$ were set to one (zero) for subdomain within the cloak (surrounding region).

To design a cloak that can simultaneously accomodate $N_l$ different loading cases, we could modify the objective function in (S7) to be

$$\min_{x^e,y^e} \left( \max_i (\mathcal{F}_i) \right), \tag{S8}$$

where $\mathcal{F}_i$ is the original objective function for the $i$th loading case. To avoid the challenges imposed by the mixed-variable min-max optimization, we approximate the maximization function with a p-norm-based aggregated function to obtain

$$\min_{x^e,y^e} \left( \tfrac{1}{N_l} \sum_{i=1}^{N_l} \mathcal{F}_i^p \right)^{1/p}, \tag{S9}$$

where $p > 1$ is a constant to control the precision of approximation. The higher the value of $p$, the more precise the approximation will be. When $p \to \infty$, (S9) will simply reduce to (S8). However, a large $p$ will make the optimization process difficult to converge. Therefore, based on preliminary studies, we use $p = 8$ in this study that could provide a good approximation of the original min-max optimization problem while enabling an effective optimization process. The cloaked structure in Figure **S10** is designed with the objective function in (S9), aggregating the performance for $N_l = 8$ different loading angles, i.e., $\boldsymbol{\theta} = \mathbf{0°, 45°, 90°}, 135°, 180°, 225°, 270°, 315°$. The resulting cloaking structure could achieve good cloaking performance for any given loading angles, as shown in Figures S10 and **S11**.

**Aperiodic Tiling Optimization**

After obtaining the optimized topology and property distribution within the cloak, for each subdomain in the cloak, a set of $N_c$ candidate unit cells ($N_c = 15$ in this study) with properties closest to the target are selected. The best unit cell should then be selected from each candidate set to assemble a mechanical cloak with compatible neighboring unit cells. To achieve this, we view the assembled metamaterial structure as a grid-like graph model shown in Figure **S4a**. Each node in the graph corresponds to a subdomain in the assembled structure with an edge connecting neighboring unit cells. For the *i*th node, we associate a label $l_i \in \{1, \cdots, N_c\}$ to represent the index of the selected unit cell in the size-$N_c$ candidate set. As shown in Figure S4b, a nodal energy $\theta_i$ is defined for the *i*th node in the graph, representing the infinite norm of the vector in the property space connecting the target and real properties of the selected unit cell $l_i$. To measure the compatibility between neighboring unit cells, the edge connecting the *i*th and *j*th nodes is associated with a nodal energy $\theta_{ij}$ defined as

$$\theta_{ij}(l_i, l_j) = \theta_{ij}^g(l_i, l_j) + w \cdot \theta_{ij}^m(l_i, l_j), \tag{S10}$$

where $\theta_{ij}^g$ is the geometrical nodal energy defined in Figure S4c as the ratio of incompatible binary elements to all solid elements on the shared boundary as a measure for the geometrical difference, $\theta_{ij}^m$ is the mechanical nodal energy, as shown in Figure S4d, defined as the relative sum of stress difference on the shared boundary under the unit strain field to measure the mechanical incompatibility, and $w$ is given constant weight. In this study, we set $w = 0.2$ but use larger weight $w = 5$ for edges across the cloaking boundary for better compatibility. With this graph model, the search for the optimal tilling of unit cells in an assembled structure is equivalent to the search for the optimal label for each node in the graph to minimize the sum of nodal and edge energies of the whole graph. It is also called the energy minimization problem on a grid-like Markov random field (MRF) and can be parallelly and efficiently solved by the dual-decomposition method.(7, 8) The optimization process will terminate when the number of iterations reaches 5000 or all subproblems agree on the nodal labeling in the dual-decomposition method. After obtaining the optimized labeling, optimal unit cells are fetched from candidate sets and mapped to the corresponding subdomain to assemble the cloaked structure.

**Calculated Displacement Fields for Different Structures**

**Figure S5** through **S9**, Figure S11 and **S12** show the calculated displacement fields for different structures as illustrated in the main text, including three examples for different boundary conditions (Figure S5~S8), one example for the cloak with optimized topology, and one example for a different base cell (Figure S9), four examples for different shapes of the void (Figure S12). **Video S1** includes animations of all the reference, voided, and cloaked structures for a circular void and their displacement fields under different boundary conditions. **Video S2** includes animations of the reference, voided, and cloaked structures for a circular void with a different base cell as in Figure S9. **Video S3** includes animations of all the reference, voided, and cloaked structures for voids of various shapes and their displacement fields under the same boundary condition. **Video S4** includes animations of the reference, voided, and cloaked structures for the Mickey-shaped void with optimized cloak region and their displacement fields. **Video S5** includes animations of the reference, voided, and cloaked structures for the butterfly-shaped voids.

**Details about experiments**

**Fabrication.** We fabricated all the structures using a commercial Stratasys Objet 500 Connex 3D printer. The maximum dimension of all the structures is $150 \times 150 \times 10$ mm. We used digital material DM8530 which is a relatively soft material obtained as a mixture of Verowhite (stiff) and Tangoblack (soft) materials. The material properties (Young's modulus $E = 1.20$ GPa and Poisson's ratio $\nu = 0.35$) were experimentally measured following ASTM D638-14 standard test method and the same values are used for finite element analysis in unit cell database construction, optimization, and numerical performance evaluation.

**Experimental Setup and Testing.** We subjected the 3D printed structures to compression using a universal testing machine, Instron E3000, mounted with a 5 kN load cell (**Figure S10**). We applied a compression displacement of 1 mm at the top boundary at a rate of 0.05 mm/s. The final state results in a boundary condition equivalent to the Displacement-Free boundary condition as shown in Figure 3a. In order to apply compression while allowing the lateral expansion, the 3D printed structures were placed into two T-slots made out of Aluminum. The structures sat into these slots while undergoing compression. The slots have a length of 155 mm inside and a width of 15 mm to leave space for free lateral expansion. We lubricated the slots with silicone lubricant to reduce friction between the 3D printed material and aluminum slots. We made sure that the slots were parallel before loading, by bringing them close to each other before the slots were tightened to the Instron grips. We also centered the slots by making appropriate markings on the slots relative to the grips. All these measures ensure that the displacement is uniformly applied over the top boundary. We observed that the load-displacement curves of all the structures to be linear validating our linear elasticity assumption.

The 3D printed structures were spray-painted white using regular off-the-shelf white paint and dried for about 30 minutes. Since the minimum feature sizes are about 0.5 mm, speckles have to be even finer. In order to achieve finer black speckles, an airbrush was used. Airbrush was kept at a distance of about 3 cm from the surface of printed structures and sprayed at an angle of about 45° until sufficient speckle density is obtained.

A Nikon D750 Camera attached to a Nikkor 200 mm f/4D IF-ED lens was used to capture images while loading. The camera has a resolution of $6016 \times 4016$ pixels. The lens was mounted onto the tripod instead of the camera to suppress noise arising due to external vibrations which will be apparent if the camera is placed on the tripod. A ring light was placed between the printed structure and camera to sufficiently illuminate the fine features on the surface of the printed structure and to obtain uniform lighting. Interval mode is used to capture images at a frequency of 1 frame per second. The normality of the lens with respect to the surface of the printed structure is achieved by fine adjustments on the tripod. Manual mode was used at an exposure rate of 1/640 sec, at an ISO setting of 1250 and an aperture setting of F8. The center of the tripod was placed at about a distance of 85 cm from the specimen. Since the camera and the Instron cannot be triggered at the same time, a few extra pictures were captured before and after the loading. In other words, Instron is triggered after the camera starts recording the pictures. Whenever the printed structure is replaced, the focus was adjusted while keeping the rest of the setup untouched to achieve consistency. With the current setup, we observe about 53 pixels per 1 mm.

**Digital Image Correlation.** The captured pictures were post-processed using a global DIC code in MATLAB designed specifically for two-dimensional materials with microstructures. (9) Initially, a mask was created on a reference image (at zero displacement) to subtract the background and create a mesh enclosing the region of interest. The obtained mesh is used to perform global DIC analysis and obtain full-field displacements. (**Figure S11**). In order to compare displacement fields of different designs quantitatively, it would be hard to measure at each mesh point since the mesh for different designs is not exactly the same. Hence, we average the displacements of nodes in a neighborhood of the lattice point intersection and use it for quantitative evaluation.

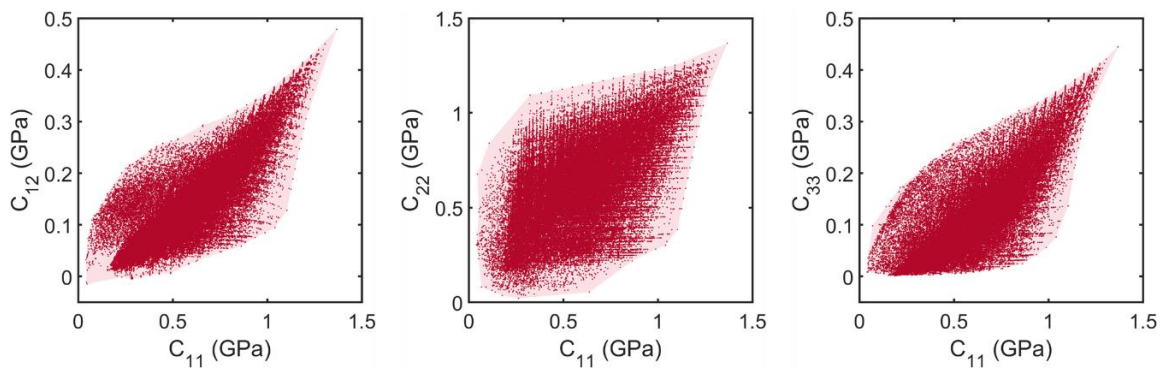

**Figure. S1.** Property space of the generated unit cell database, with shaded regions indicating the boundary.

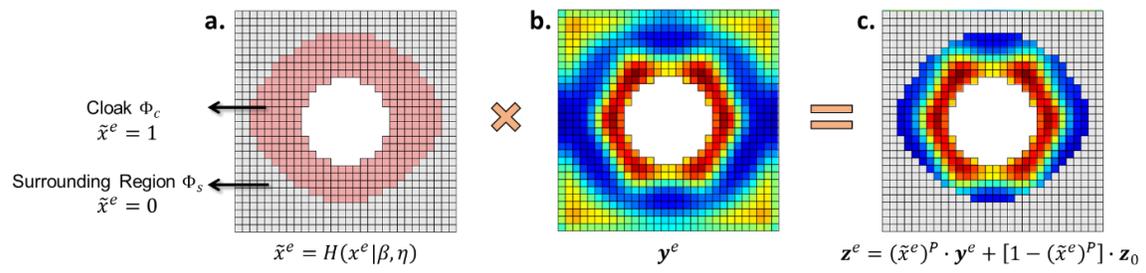

**Figure. S2.** Variable definition in the topology and properties optimization, a. projected topology design variable, b. property design variable, c. normalized property vector

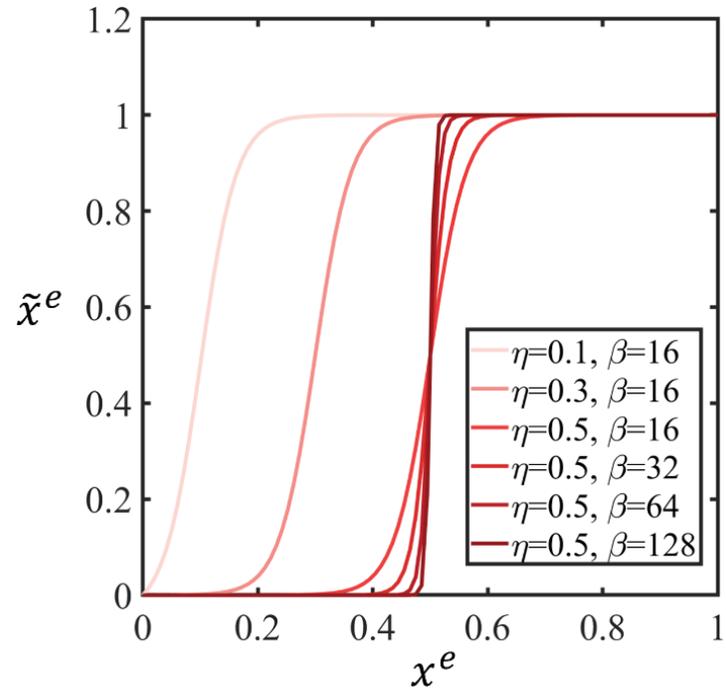

**Figure. S3.** Projection scheme in Equation S3 with different combinations of parameters $\beta$ and $\eta$.

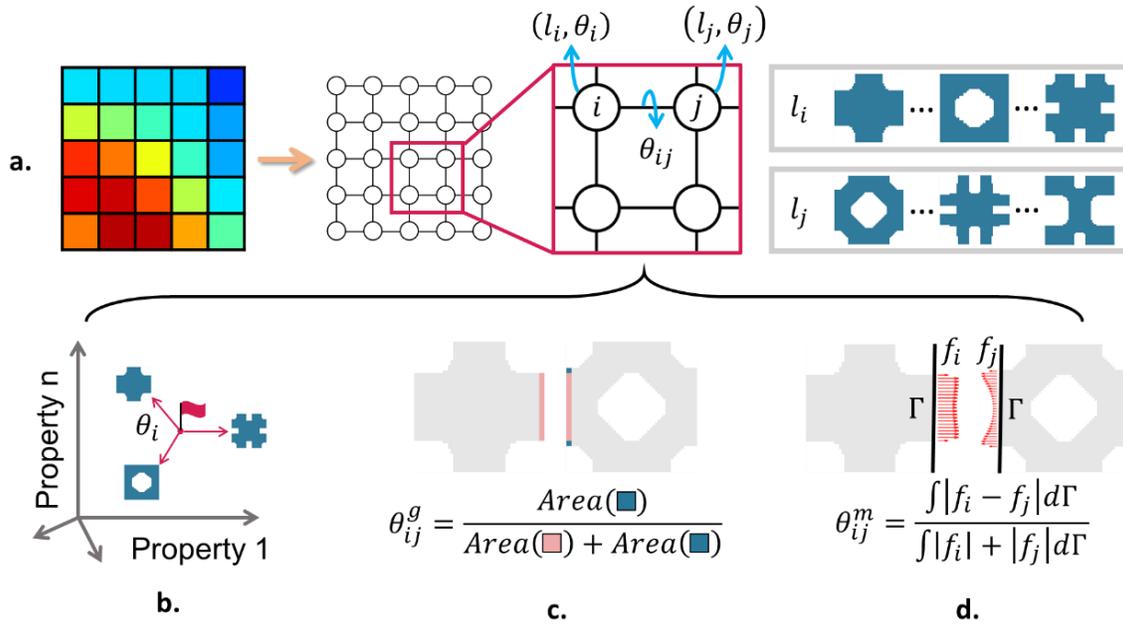

**Figure. S4.** Illustration of the tiling optimization process, a. transforming the tiling problem into an equivalent energy minimization problem on a grid-like graph, b. nodal energy, b. geometrical edge energy, c. mechanical edge energy.

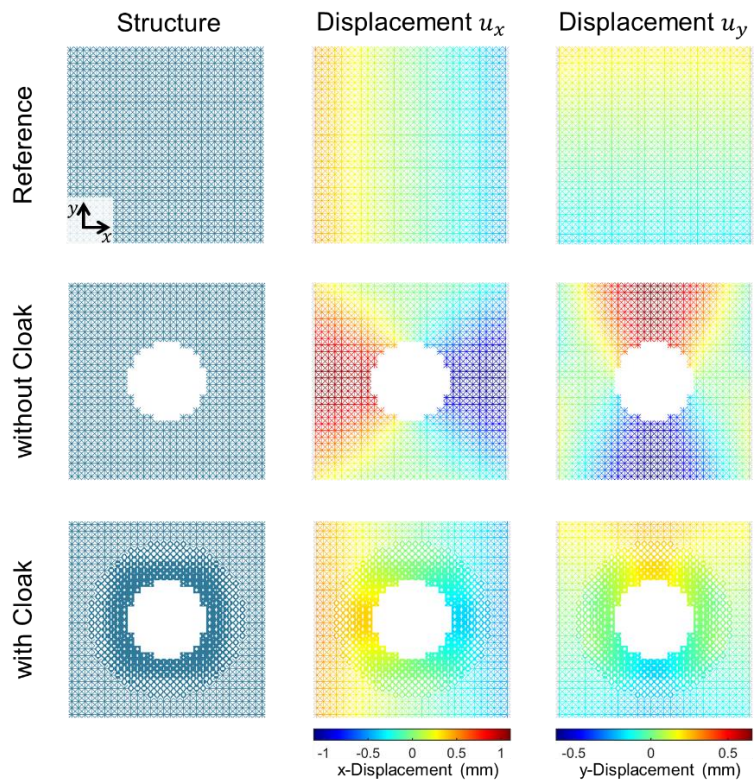

**Figure. S5.** Structures and calculated displacement fields as in Figure 2, but under the pressure-free boundary condition. Constant pressure is imposed on the left and right boundaries for each structure. The cloaked structure corresponds to the structure shown in Figure 3c.

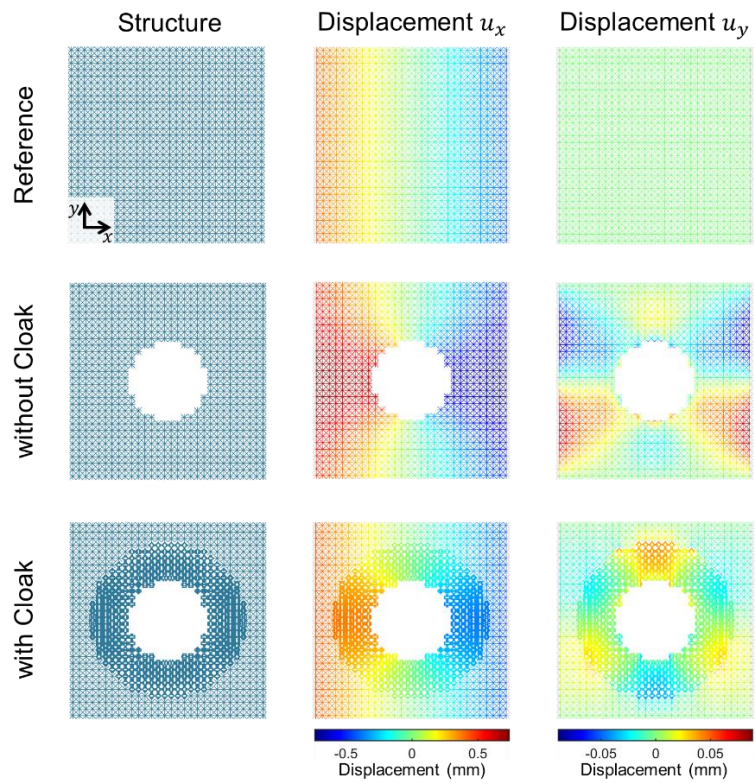

**Figure. S6.** Structures and calculated displacement fields as in Figure 2, but under the pressure-sliding boundary condition. Constant pressure is imposed on the left and right boundaries while sliding boundary condition is imposed on the top and bottom for each structure. The cloaked structure corresponds to the structure shown in Figure 3d.

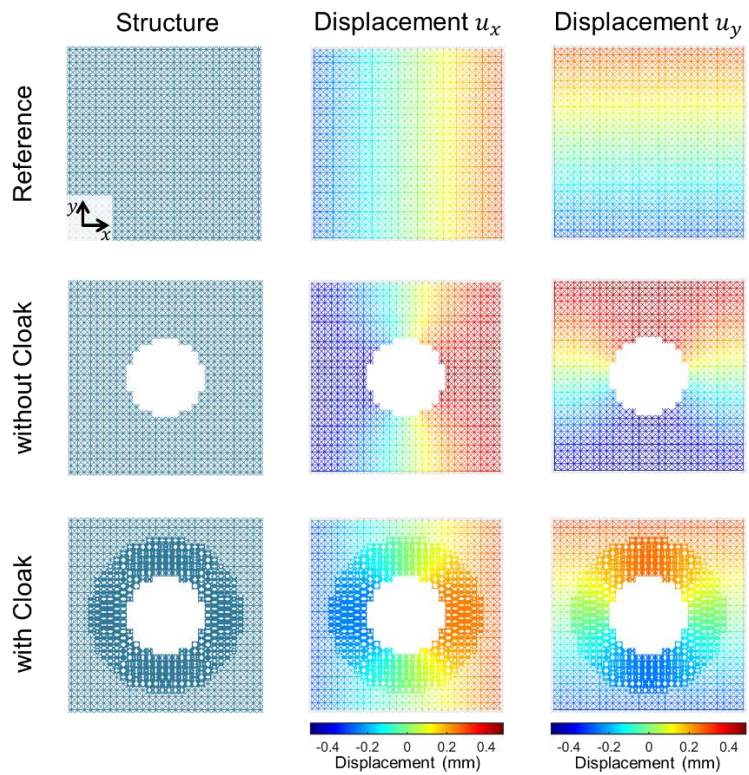

**Figure. S7.** Structures and calculated displacement fields as in Figure 2, but under the dilating boundary condition. Constant uniformly distributed stretching forces are exerted from all four boundaries for each structure. The cloaked structure corresponds to the structure shown in Figure 3e.

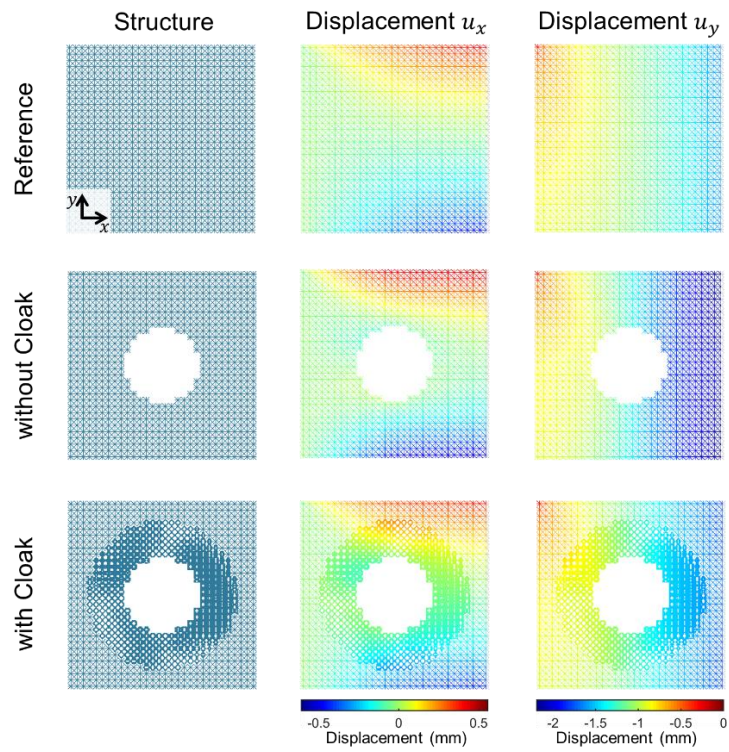

**Figure. S8.** Structures and calculated displacement fields as in Figure 2, but under the shearing boundary condition. Constant uniformly distributed shearing forces are imposed on the right boundary while sliding boundary condition is imposed on the left for each structure. The top-left corner is fixed. The cloaked structure corresponds to the structure shown in Figure 3f.

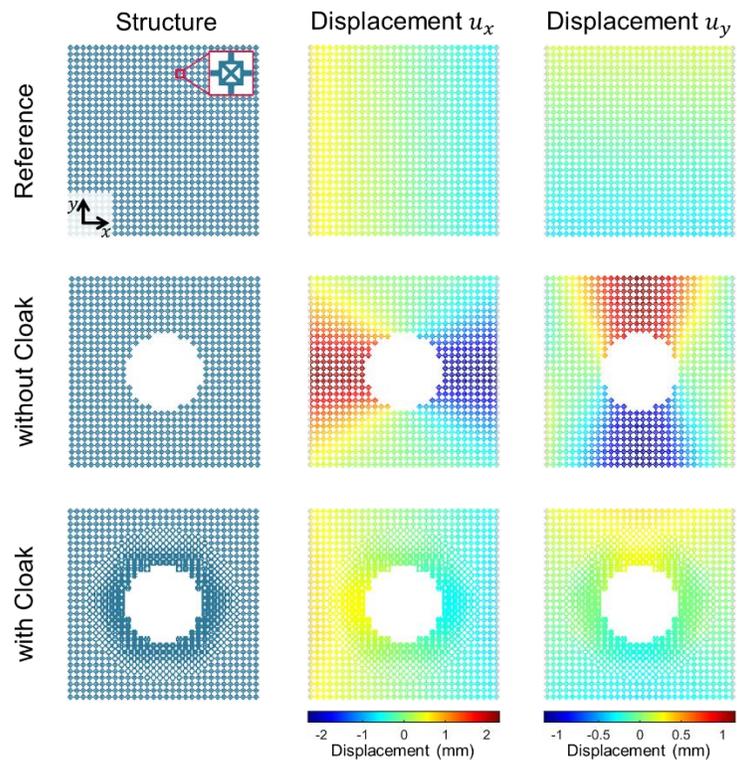

**Figure. S9.** Structures and calculated displacement fields as in Figure S5, but with a different base cell. An enlarged image for the base cell is shown in the reference structure. The cloaked structure corresponds to the structure shown in Figure 3g.

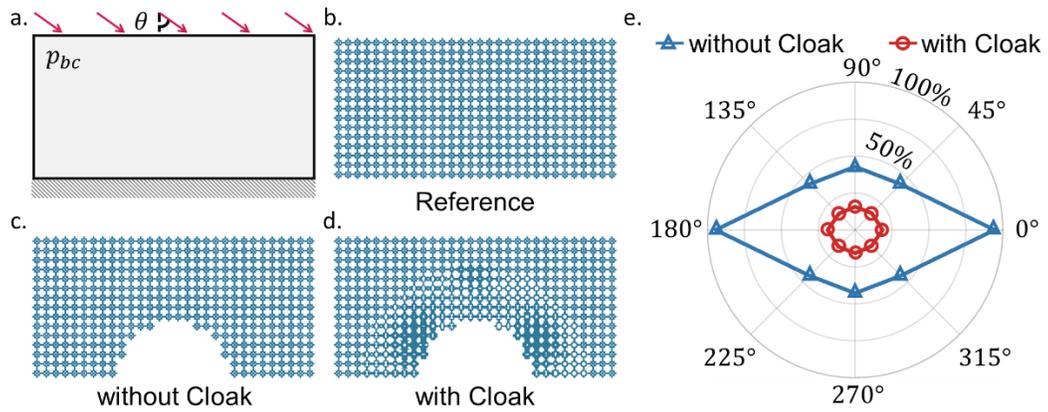

**Figure. S10.** Boundary condition, structures, and relative difference values of a cloak devised to simultaneously accommodate all possible loading angles, (a) boundary condition definition, a constant distributed loading $p_{bc} = 95.67 \text{ KPa}$ is imposed on the top while the bottom is fixed, the pressure has the same magnitude as in Figure 2 but with a varying loading angle $\theta \in [0°, 360°)$, (b) reference structure composed of base unit cells as in Figure S9, (c) voided structure, (d) cloaked structure, (e) polar plot for relative difference values of voided and cloaked structures given different loading angles $\theta$, with the relative difference values as the radius axis. The markers show the discrete loading angles used for obtaining the aggregated objective function in equation (S9).

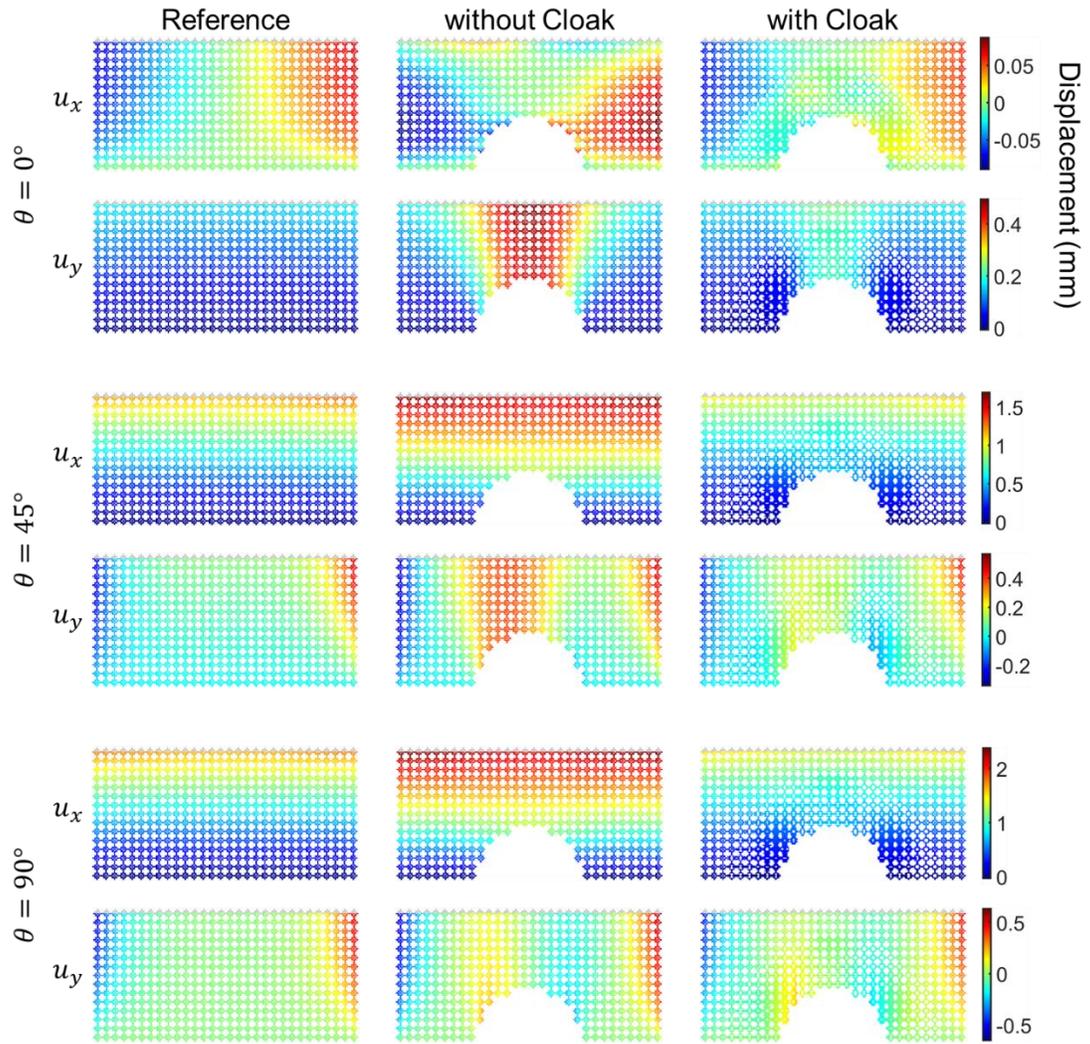

**Figure. S11.** Calculated displacement fields for structures in Figure S10, given three different loading angles $\theta = 0°, 45°, 90°$.

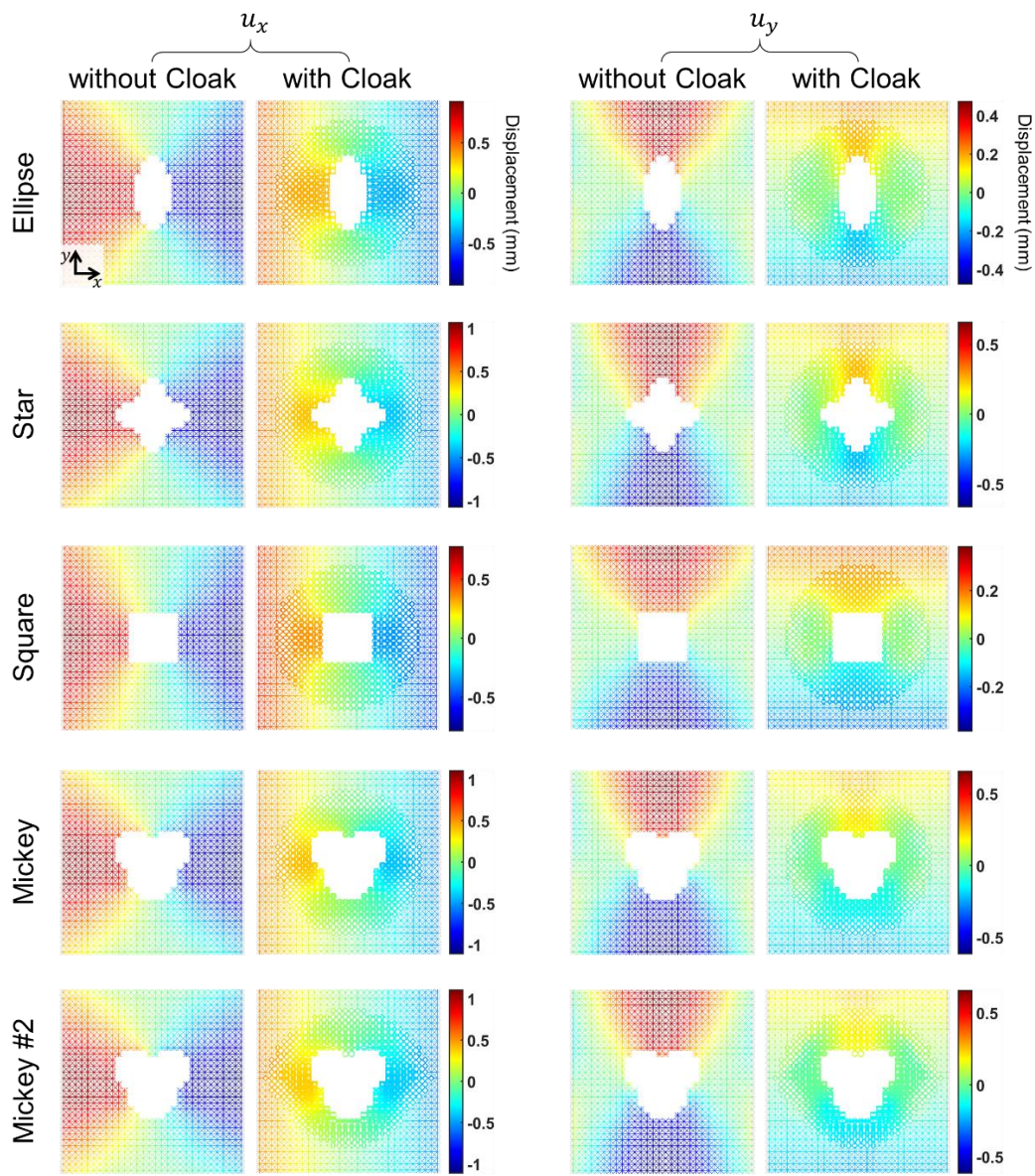

**Figure. S12.** Calculated displacement fields for structures with and without the designed cloak, given the same pressure-free boundary condition as in Figure S5 but with different shapes of the void. The cloaked structure in the last row (Mickey #2) has a designed cloak with optimized topology while the others have predefined and fixed topology for the cloak as in Figure 2c. These cloaked structures correspond to the structures shown in Figure 3h~3l.

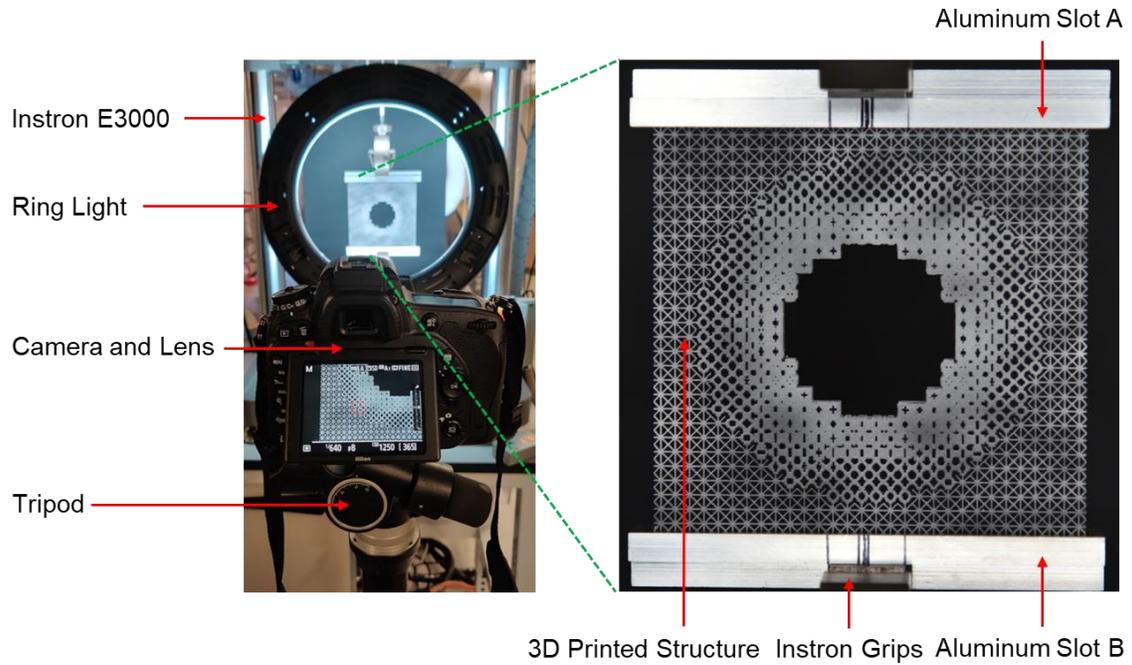

**Figure. S13.** Experimental setup consisting of 3d printed structures being compressed between two aluminum slots by universal testing machine (Instron E3000). A series of images are captured using a camera mounted on a tripod while a ring light is used to obtain uniform lighting over the structure.

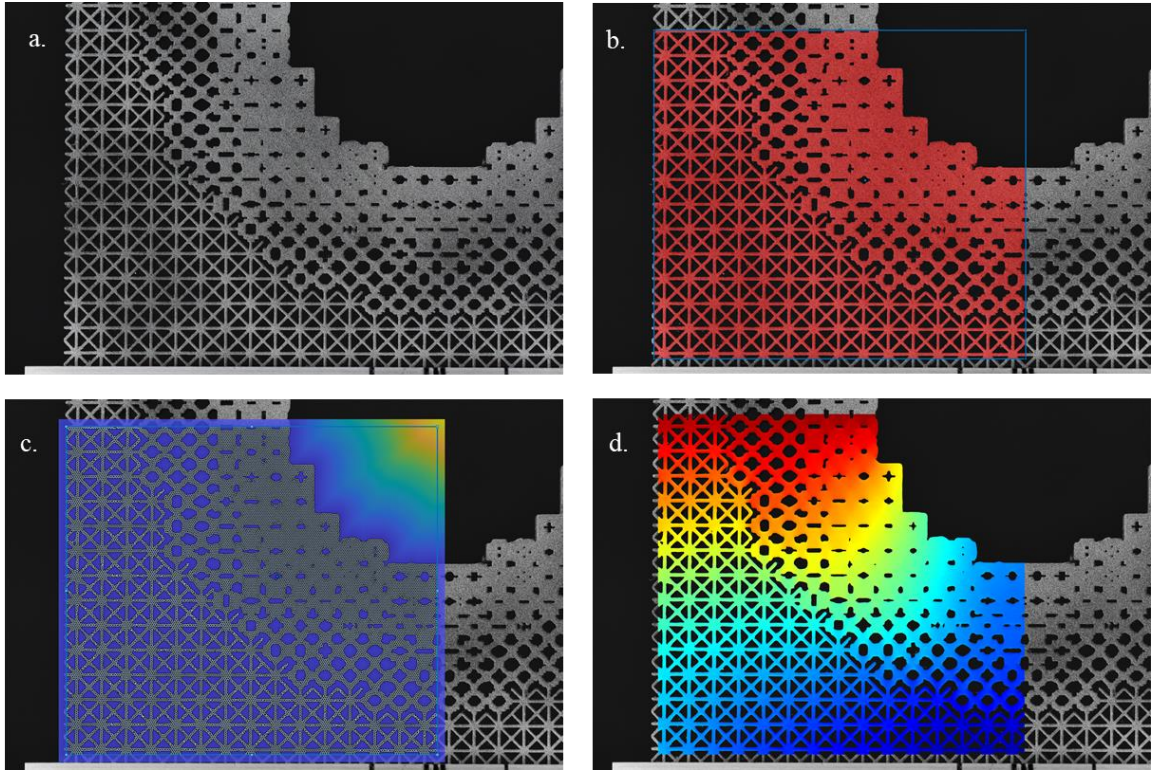

**Figure. S14.** Digital Image Correlation: a.) Speckled structure as observed in the camera. b.) Generation of a mask to define ROI c.) Triangular mesh generated in the ROI excluding the areas where there is no material d.) Full- displacement fields obtained from the DIC code

**Movie S1 (separate file).** Deformation of structures under different loading conditions.

**Movie S2 (separate file).** Deformation of structures with a different base cell.

**Movie S3 (separate file).** Deformation of structures with different shapes of the void.

**Movie S4 (separate file).** Deformation of structures with adaptive cloaking region.

**Movie S5 (separate file).** Deformation of structures with butterfly-shaped voids.

**SI References**

1. L. Wang, Y.-C. Chan, Z. Liu, P. Zhu, W. Chen, Data-driven metamaterial design with Laplace-Beltrami spectrum as "shape-DNA". Structural and Multidisciplinary Optimization, 1-16 (2020).

2. L. Wang et al., Deep generative modeling for mechanistic-based learning and design of metamaterial systems. Computer Methods in Applied Mechanics and Engineering 372, 113377 (2020).

3. L. Xia, P. Breitkopf, Design of materials using topology optimization and energy-based homogenization approach in Matlab. Structural and multidisciplinary optimization 52, 1229-1241 (2015).

4. Z. Hashin, Analysis of Composite Materials—A Survey. Journal of Applied Mechanics 50, 481 (1983).

5. X. Qian, Undercut and overhang angle control in topology optimization: a density gradient based integral approach. International Journal for Numerical Methods in Engineering 111, 247-272 (2017).

6. K. Svanberg, The method of moving asymptotes—a new method for structural optimization. International journal for numerical methods in engineering 24, 359-373 (1987).

7. C. Wang, N. Komodakis, N. Paragios, Markov random field modeling, inference & learning in computer vision & image understanding: A survey. Computer Vision and Image Understanding 117, 1610-1627 (2013).

8. N. Komodakis, N. Paragios, G. Tziritas, MRF energy minimization and beyond via dual decomposition. IEEE transactions on pattern analysis and machine intelligence 33, 531-552 (2010).

9. F. Agnelli, P. Margerit, P. Celli, C. Daraio, A. Constantinescu, Systematic two-scale image analysis of extreme deformations in soft architectured sheets. International Journal of Mechanical Sciences 194, 106205 (2021).